\documentstyle[12pt,feyn]{article}
\catcode`@=11
\def\seceqaa{\@addtoreset{equation}{section}
           \def\theequation{A\arabic{equation}}}
\def\seceqbb{\@addtoreset{equation}{section}
           \def\theequation{B\arabic{equation}}}
\def\seceqcc{\@addtoreset{equation}{section}
           \def\theequation{C\arabic{equation}}}
\catcode`@=12
\begin{document}
\title{Absence of Nonlocal Counter-terms in the Gauge Boson Propagator in Axial -type
Gauges}
\author{{Satish. D. Joglekar}\thanks{e-mail:sdj@iitk.ac.in}\\
Department of Physics, Indian Institute of Technology\\
 Kanpur 208 016, UP, India\\
{A. Misra} \thanks{e-mail:aalok@iitk.ac.in}\\
Institute of Physics, Sachivalaya Marg\\
Bhubaneswar 751 005, Orissa, India}
\maketitle

\begin{abstract}
We study the two-point function for the gauge boson in the axial-type gauges.We
use the exact treatment of the axial gauges recently proposed that is 
intrinsically
compatible with the Lorentz type gauges in the path-integral formulation and
has been arrived at from this connection and which is a ``one-vector'' 
treatment.
We find that in this treatment, we can evaluate the two-point functions without
imposing any additional interpretation on the axial gauge 
1/\( (\eta  \).q)\( ^{q} \)
type poles.The calculations are as easy as the other treatments based on other
known prescriptions.Unlike the ``uniform-prescription'' /L-M
prescription, we
note,here, the absence of any non-local divergences in the
 2-point proper vertex. We correlate our calculation with 
that for the Cauchy Principal Value prescription
and find from this comparison that the 2-point proper vertex differs from the
CPV calculation only by finite terms.For simplicity of treatment,
the divergences
have been calculated here with \( \eta ^{2}>0 \) and these have a smooth light
cone limit.
\end{abstract}

\section{Introduction}

The calculations in gauge theories have been done primarily 
using the Lorentz-type
{[}including the R\( _{\xi } \)-gauges{]} and the axial type gauges {[}1{]}.
The Lorentz type gauges have the several desirable 
properties of Lorentz- covariance
,ease of calculations and the availability of a gauge parameter to verify the
gauge-independence. They , however, are burdened with having to include the
diagrams involved with the Faddeev-Popov ghosts.As a result of this, another
set of gauges ,the axial-type gauges{[} these include the light-cone {[}LCG{]}
and the temporal gauges{]}, 
have also been employed in the Standard Model {[}SM{]}
calculations.These gauges purport to have decoupling of ghosts {[}2,3,4{]} and
consequently require a fewer set of diagrams.The disadvantages these gauges
suffer from are the lack of manifest covariance, extra counter-terms arising
from an additional vector \( \eta  \) \{ or \emph{two} additional vectors in
case of the Leibbrandt-Mandelstam/Uniform prescriptions {[}2,3{]}\} and the
problem of how to correctly deal with the spurious singularities of the form
1/\( (\eta  \).q)\( ^{q} \).Doubts have also been expressed {[}5{]} about
the exceptional advantages of the axial-type gauges;nonetheless they have been
found useful in practice. The crucial problem of 1/\( (\eta  \).q)\( ^{q} \)
-type singularities has been widely addressed 
over several decades {[}2,3,6-8{]}. The
early use of principal value prescription came 
under cloud on account of several
difficulties encountered with it. It was found that the PVP could not be used
for the light-cone gauges for several reasons {[}3,4{]}.Further, it was shown
that the PVP does not give the correct behavior 
for the Wilson loop to O{[}g\(^{4}] \)
compatible with the calculation in Feynman gauge{[}9{]}. 
To overcome these difficulties,
the Leibbrandt-Mandelstam prescription {[}2,3,4{]} was proposed and applied
to the light-cone gauges.It was further generalized to the other axial gauges
through what is called the uniform prescription {[}2,3{]}. One of the drawbacks
of the L-M/Uniform prescriptions has been the presence of the 
non-local counterterms
in proper vertices to all orders. Another drawback is, 
of course,that the presence
of two vectors leads to a larger set of possible divergences. 
Bassetto and coworkers
have developed the techniques for dealing with non-local divergences {[}2,3{]}.
Several other ways of dealing with the 
1/\( (\eta  \).q)\( ^{q} \)-type singularities
have also been proposed {[}6-8{]}.

A new way of attacking the problem of 
1/\( (\eta  \).q)\( ^{q} \) -type singularities
in axial type-gauges was formulated recently.This was based on establishing
the relation between the Green's functions in the Lorentz and the axial-type
gauges in the path-integral formulation.This relation was, in turn, based on
what was called the finite field-dependent 
BRS {[}FFBRS{]} transformation {[}10{]}.
According to this view-point, the crucial 
question is how to develop an \emph{intrinsically
well-defined} path-integral treatment in any other gauge that is compatible
to that in the Lorentz gauges \emph{by construction} that, we understand have
no analogous difficulties. Such a well-defined treatment 
for defining the path-integral
in axial-type gauges {[}and which is also applicable to host of other gauges
{[}11{]}{]}has been established using the 
FFBRS transformation {[}12,13,7,8{]}.It
is expected that this path-integral so constructed should provide answers to
all the questions regarding the various difficulties 
associated with the variety of the non-covariant gauges 
{[}e.g. axial, light-cone, temporal
{[}2,3{]}, Coulomb {[}14{]} that are thought to arise from the
 ill-definedness of the naive treatment.
In other words,as we have emphasized, we expect that the way to deal with these
difficulties does not require an ad hoc augmentation of rules as to how the
diagrams are calculated, but is already contained in the 
process outlined earlier
in references {[}11,12,13{]}.The above expectation ,based on rigorous formal
arguments, derives concrete support from several works
{[}7,8,15{]}. For example following the above outlook {[}16{]}, we 
have established an effective treatment
of the axial gauge propagator {[}7,8{]}.We have also shown that the Wilson loop
for axial and Lorentz gauges has the same value to O{[}g\( ^{4}] \) for a wide
class of loops {[}15{]}.

With this in mind ,we perform a simple one-loop calculation 
of the 2-point proper
vertex in pure gauge theories.We have several motivations to perform
this simple calculation. According to the view-point 
mentioned earlier, our calculation does not have any arbitrariness 
associated with it and is in a formalism intrinsically
compatible with the Lorentz gauges.Our calculation involves 
evaluations of terms
having contour integrals and those with effective \( \delta  \)-function terms.
We exhibit that our calculations are no more cumbersome than those with other
prescriptions and that the simplifications that are normally associated with
dimensional regularization {[}DR{]} continue to hold in our treatment. We find
it easy and convenient to correlate our calculations with those with
PVP \footnote{
We do not necessarily imply correctness or otherwise of the use of PVP by this
comparison.} and point out the additional contributions.
We show how these can be dealt with and show that these are in fact
finite. Our method ,\emph{without any interpretation
given to Feynman integrals,} allows one to verify that the ghost diagram does
not contribute to the \emph{proper} two point function.

We summarize the plan of the paper.In section 2, we introduce the notations
and the past results obtained from the FFBRS treatment
{[}11,12,7,8{]}. In section 3, we give the procedure for evaluation of
the three diagrams, both the contour
integral-type contribution and the effective term-type contributions.We show
that the evaluations are in no way more involved as compared to those with say,
CPV or L-M prescriptions. We emphasize that we, in no way, need to 
\emph{interpret}
any of the term in the Feynman
diagrams. We show here ,in particular, that the
results that we normally expect for the tadpole 
diagram in the dimensional regularization
and from the ghost-decoupling for the ghost diagram hold in this
treatment  also. Again, the ghost diagram does not contribute
\emph{without} requiring any assumptions
made about 1/\( \eta .q \) type terms.In section 4, we summarize our results
and indicate heuristically the direction one can take to tackle the question
of locality of divergences in higher orders and for other n-point functions.
Appendix B deals with the ghost diagram.

\section{Preliminaries}

We work in the axial-type gauges with a gauge parameter $\lambda$
and
\begin{equation}
\label{eq:2.1}
S^A_{\rm eff} = -{1\over
  2\lambda}\int d^4x(\eta\cdot A)^2
\end{equation}
and regard $\eta\cdot A=0$ as the $\lambda\rightarrow0$ limit of
the above family of gauges. We also have the ghost action:
\begin{equation}
\label{eq:2.2}
S^A_{\rm gh}=-\int d^4x{\bar c}\eta^\mu D_\mu c.
\end{equation}
In references {]}7{]} and {]}8{]}, we
derive the treatment of the axial
 gauge poles by
connecting the axial gauges to the Lorentz gauges
established in earlier works {]}12{]} and {]}14{]}.. We found 
the exact axial gauge propagator in this way and a much simpler
effective treatment for it. The propagator reads 
\begin{equation}
\label{eq:freslt1}
\tilde G^{0 A}_{\mu\nu}=\tilde G^{0 L}_{\mu\nu}\
+\biggl[\biggl(k_\mu k_\nu\Sigma_1+\eta_\mu k_\nu\Sigma_2\biggr)ln\Sigma_3
+(k\to -k;\mu\leftrightarrow\nu)\biggr]
\end{equation}
where
\begin{eqnarray}
\label{eq:freslt2}
& & \Sigma_1\equiv {-(k^2-i\eta\cdot k)\biggl({\eta\cdot k+i\eta^2
\over{k^2-i\eta\cdot k}}+i\lambda
-{(1-\lambda)\eta\cdot k\over{k^2+i\epsilon}}\biggr)
\over{\epsilon\Sigma}}\nonumber\\
& & \Sigma_2\equiv {-(k^2-i\eta\cdot k)\biggl(
-\biggl[{k^2+i\eta\cdot k\over{k^2-i\eta\cdot k}}
\biggr]+1-{i\epsilon(1-\lambda)\over{k^2+i\epsilon}}\biggr)
\over{\epsilon\Sigma}}\nonumber\\
\nonumber\\
& & \Sigma_3\equiv{-i(\eta\cdot k+\epsilon)(k^2+i\epsilon\lambda)
\over{(k^2+i\epsilon)\biggl(-i\epsilon\lambda-\sqrt{k^4-(k^2+i\epsilon\lambda)
\biggl[k^2+{(\eta\cdot k)^2+i\epsilon\eta^2\over{k^2+i\epsilon}}\biggr]}\biggr)}},\nonumber\\
& & {\rm and}\nonumber\\
& & \Sigma\equiv\biggl[(1-\lambda)[(\eta\cdot k)^2+2ik^2\eta\cdot k]
+i\epsilon k^2(1-2\lambda)+\lambda(k^2+i\epsilon)^2+\eta^2(k^2+i\epsilon)\biggr].
\nonumber\\
& &
\end{eqnarray}

Despite the formidable appearance of the above, a much simpler
effective treatment was also established
in {[}7{[}. It was shown that coordinate space propagator 
$D_{\mu\nu}(x,y)$ should be evaluated as shown below. We
shall take $\eta_0\neq0$ and in fact let $\eta_0=1$.
\footnote{We shall find it convenient to deal with the
  case $\eta^2>0$ as this makes the treatment simpler.}Below, $C$
represents a contour from $(-\infty,\infty)$ along the real
$k_0$-axis except a semicircular dip of
radius $\alpha\sqrt{\epsilon}(\alpha>>1$) below $\eta\cdot
k=0$.
We have 
\begin{equation}
\label{eq:2.5}
D_{\mu\nu}(x,y)=\int d^3k\int_C
dk^0e^{-ik\cdot(x-y)}D^0_{\mu\nu}(k)
+\int d^3k\int_{-\infty}^{\infty}
e^{-ik\cdot(x-y)}
D^{\rm extra}_{\mu\nu}(k),
\end{equation}
with $D^0$ as the usual axial propagator away from $\eta\cdot
k=0$ axis, viz
\begin{equation}
\label{eq:2.6}
D^0_{\mu\nu}(k)_
=-{1\over{k^2+i\epsilon}}\bigl(g_{\mu\nu}-{k_\mu\eta_\nu+
k_\nu\eta_\mu\over\eta\cdot k}+k_\mu k_\nu{\lambda
  k^2+\eta^2\over{(\eta\cdot k)^2}}\biggr),
\end{equation}
and {[}7{[}
\begin{eqnarray}
\label{eq:Deff}
& & D^{\rm extra}_{\mu\nu}(k)=
\delta\biggl(k^0
-{1\over2}\sqrt{{\epsilon\eta^2\over i}}-\vec\eta\cdot\vec k\biggr)
\biggl(k_\mu k_\nu
\biggl[-i\sqrt{{i\eta^2\over\epsilon}}a_{1(0)}
-{i\eta^2\over 2}a_{1(1)}\biggr]\nonumber\\
& & +\eta_\mu k_\nu\biggl[-{i\eta^2\over 2}a_{2(1)}\biggr]
+\eta_\nu k_\mu\biggl[-{i\eta^2\over 2}a_{3(1)}\biggr]\biggr)\nonumber\\
& & =\delta\biggl(k^0
-{1\over2}\sqrt{{\epsilon\eta^2\over i}}-\vec\eta\cdot\vec k\biggr)\nonumber\\
& & \times
\biggl( -{2\pi\eta^2\over{(\eta^2+i\epsilon)[(\vec\eta\cdot\vec
    k)^2-\vec k^2]}}
\biggl[ i\sqrt{{i\eta^2\over\epsilon}}+{\eta^2\over
  \eta^2+i\epsilon}\biggr]k_\mu
k_\nu-{2\pi\over{(\eta^2+i\epsilon)
[(\vec\eta\cdot\vec k)^2-\vec k^2]}}(\eta_\mu k_\nu+k_\nu
\eta_\mu)
\biggr).\nonumber\\
& & 
\end{eqnarray}
We \footnote{We note that for
  $\eta^2>0$, the denominator
$(\vec\eta\cdot\vec k)^2-\vec k^2$ never vanishes and we
can simplify by dropping $\epsilon$ that went along with it.}
shall find it beneficial to compare our calculations
with the
Cauchy Principle Value prescription (CPV)
calculations for which
\begin{equation}
\label{eq:CPV1}
{1\over{(\eta\cdot k)^\beta}}\rightarrow{1\over2}
\lim_{\mu\rightarrow0}\biggl[{1\over(\eta\cdot 
k+i\mu)^\beta}+{1\over{(\eta\cdot k-i\mu)^\beta}}\biggr].
\end{equation}

\section{One-Loop Calculations}

In this section, we shall pursue the one-loop calculations
with the present formalism. We illustrate how treatment of the Feynman
integrals involving contours of the type as 
in (\ref{eq:2.5}), and that of the extra effective terms. We, in
particular, emphasize that
our calculations are no way more laborious, despite its new
appearance, than those performed 
with prescriptions such as CPV and Uniform  prescriptions
{]}2{]}, {]}3{]}. We further wish to emphasize that our calculations 
are done from
first principles and are compatible with those
of Lorentz-type gauges {]}13{]}, {]}11{]}.by construction
and
involve no arbitrariness like the other 
prescriptions . Our method, in
addition,
has only one vector $\eta$ unlike the uniform prescription and 
thus does not increase the number of possible
counterterms.. Moreover,
we find, as shown below, that we do not encounter non-local
divergence (in this simple example) found with uniform/ML
prescription and we suspect this to hold
in higher order calculations. Unlike CPV, this treatment has been
shown explicitly to preserve a large class of 
Wilson lops (i.e. leads to the same value for it as in Lorentz gauges
{]}15{]}). We find that the divergences found have a
smooth limit
$\eta^2\rightarrow0$ and 
leads to no nonlocal
divergences.

We shall find it convenient to 
correlate our calculations with those done with CPV prescription. We
shall in fact show that the 
contributions, over and above CPV, both from the
contour
integral and the effective terms are
finite. We shall successively deal
with the three diagrams in 
Fig 1. We shall show that in our effective treatment also, the
tadpole diagram
as well as the ghost diagrams vanish exactly 
(i.e. without an interpretation given to the propagator near
$\eta\cdot
k=0$).

As for the tadpole diagram, it is proportional to 
\begin{equation}
\label{eq:tadpole1}
\int d^4 k D^\mu_\mu(k
) = \int d^3k
\int_C d k_0 D^{0\ \mu}_\mu (k)+\int d^3k\int dk_0
D^{\rm  extra\ \mu}_\mu(k).
\end{equation}
From
(\ref{eq:2.6}), 
\begin{eqnarray}
\label{eq:3.2}
& & D^{0\mu}_\mu(k)=-{2\over
  k^2+i\epsilon}
-{\lambda k^2+\eta^2\over(\eta\cdot k)^2}
+O(\epsilon)\nonumber\\
& & = -{2\over k^2+i\epsilon}-\lambda-{\lambda\eta\
\cdot k} -{\lambda[(\vec\eta \cdot\vec k)^2
-\vec k^2]+\eta^2\over(\eta\cdot k)^2}.
\end{eqnarray}
We note that the term $(-\lambda)$ can be dropped ($\int
d^nk=0$). The contour integral over
$C$ can be obtained by closing it
below for the first and the last
terms; the last term then
does not
contribute. We obtain, using symmetric integration for the
third term:
\begin{equation}
\label{eq:3.4}
\int_C D^{0\mu}_\mu(k)={2\pi i\over|\vec
  k|}
-\lambda\pi
i\vec\eta\cdot\vec
k 
\end{equation}
We, the, note that the first term in (\ref{eq:tadpole1}) vanishes noting
that in dimensional regularisation,
\begin{equation}
\label{eq:3.5}
\int{d^{n-1}k\over|\vec k|}=\int
d^{n-1}k
\vec\eta\cdot\vec k=0.
\end{equation}
We note that the second term in (\ref{eq:tadpole1}) vanishes noting from 
(\ref{eq:Deff}) that 
\begin{equation}
\label{eq:3.6}
D^{\rm extra\mu}_\mu(k)={\rm a\ \vec
  k-independent\ constant},
\end{equation}
and $\int
d^{n-1}k=0$.
Thus, the diagram vanishes here also.

We shall, next, take up diagram 1(b). We shall first establish
the
treatment for the graph in terms of a contribution over
$C$ and an effective contribution. Up to overall factors, the
graphs
is of the form,
\begin{equation}
\label{eq:3.7}
\int d^4k
V_{\mu\rho\lambda}(p,k,-k-p)V_{\nu\kappa\sigma}
(-p,k+p,-k)D^{\sigma\rho}(k)D^{\lambda\kappa}(k+p),
\end{equation}
where $D^{\sigma\rho}$ is the exact propagator found in
{[}7{]}, {[}8{]}. We shall reduce the expression
    (\ref{eq:3.7}) in terms of the contour 
integral
and an effective term. We do this in Appendix A for
$\eta\cdot
p\neq0$. As shown in Appendix A, and not
surprisingly, we
find:
\begin{eqnarray}
\label{eq:3.8}
& & \int_{-\infty}^{\infty}dk_0 VVDD=\int_{C_0}dk_0
VVD^0(k)D^0(k+p)+\int dk_0 VVD^0(k+p)D^{\rm extra}
(k)\nonumber\\
& & +\int dk_0 VVD^0(k)D^{\rm extra}(k+p).
\end{eqnarray}
Here $C_0$ is the contour running along the real
$k_0$-axis
except for two semicircular
dips of
radius  $\alpha\sqrt{\epsilon}<<|\eta\cdot
p|$
centered at $\eta\cdot
k=0$ and
$\eta\cdot(k+p)=0$.

In Appendix A, we have further established a simplification
in the first term on the right hand side of
(\ref{eq:3.8}). It is
shown there that in
evaluating 
the term, we only have to
consider integrals of the form:
\begin{equation}
\label{eq:3.9}
\int_C
d^nk{P(k,p)\over{(k^2+i\epsilon)[(k+p^2+i\epsilon]\zeta^\alpha}},\ 
0\leq\alpha\leq2,
\end{equation}
where $C$ is the contour
with only a single semicircular 
dip around $\eta\cdot k=0$, and 
$D^0(k)$ is the naive propagator of
(\ref{eq:2.6}).

We show that the divergences in the
integral
of the form (\ref{eq:3.9})
can in fact, be related to their evaluations in 
the CPV scheme. We recall the result for
\begin{eqnarray}
\label{eq:3.10}
& &
\int_{\rm CPV}
d^nk{P(k,p)\over{(k^2+i\epsilon)[(k+p^2+i\epsilon]\zeta^\alpha}}
\nonumber\\
& & =
\lim_{\mu\rightarrow 0}\biggl({1\over2}\int d^nk\int_{-\infty}^{\infty}
dk_0
{P(k,p)\over{(k^2+i\epsilon)[(k+p)^2+i\epsilon]
(\zeta-i\mu)^\alpha}}\nonumber\\
& &  +
\int d^nk\int_{-\infty}^{\infty}
dk_0{P(k,p)\over{(k^2+i\epsilon)[(k+p)^2+i\epsilon]
(\zeta+i\mu)^\alpha}}\biggr).\nonumber\\
& & 
\end{eqnarray}
In the first integral on the right hand
side
of (\ref{eq:3.10}), we can deform the contour along the real 
$k_0$-axis to $C$ (i.e. with a dip
 at $\zeta=0$) and allow
$\mu\rightarrow$. Similarly, in the second
integral, we can
deform the contour to
$C^\prime$ (with a bump above at
$\zeta=0$)
and allow
$\mu\rightarrow$. Thus, we
have
\begin{equation}
\int_{\rm
CPV}
dk_0{P(k,p)\over{(k^2+i\epsilon)[(k+p^2+i\epsilon]\zeta^\alpha}}
={1\over2}\biggl(\int_C+\int_{C^\prime}\biggr)
dk_0{P(k,p)\over{(k^2+i\epsilon)[(k+p^2+i\epsilon]\zeta^\alpha}}.
\end{equation}
We thus see that
\begin{eqnarray}
\label{eq:3.15}
& & \int_C dk_0{P(k,p)\over{(k^2+i\epsilon)[(k+p^2+i\epsilon]\zeta^\alpha}}
-\int_{\rm CPV}
k_0
{P(k,p)\over{(k^2+i\epsilon)[(k+p)^2+i\epsilon]\zeta^\alpha}}
\nonumber\\
& & ={1\over2}\oint
{P(k,p)\over{(k^2+i\epsilon)[(k+p)^2+i\epsilon]\zeta^\alpha}}
\end{eqnarray}
where
$\oint_C$ goes over a circular contour of
radius
$=\alpha
\sqrt{\epsilon}$ around
$\zeta=0$.

In evaluating the diagram 1(b), we then have to 
evaluate integrals of the form
\begin{equation}
\label{eq:3.16}
I_1=\int
d^3k\int_{C_0} dk_0
V_{\mu\rho\lambda}V_{\nu\xi\sigma}D^{0\sigma\rho}(k)
D^{0\lambda\xi}(k+p)
\end{equation}
and
\begin{equation}
\label{eq:3.17}
I_2=\int
d^3k\int_{C_0} dk_0
V_{\mu\rho\lambda}V_{\nu\xi\sigma}D^{0\sigma\rho}(k)
D^{{\rm extra}\lambda\xi}(k+p)
\end{equation}
(and an analogous term $I^\prime_2$ with 
$D^{{\rm extra}\sigma\rho}(k)D^{0\lambda\xi}(k+p)).$
In (\ref{eq:3.16}), we keep
aside  the
Feynman gauge term, as this has only a local divergence. in other
terms in (\ref{eq:3.16}), we apply the simplifications due to 
(i) tree WT-identity for $p_\mu V^\mu$ etc. and (ii)
partial fraction indicated in Appendix A (See equation
(A.5)) and arrive at integrals of the form of
(\ref{eq:3.15}). These
can be evaluated firstly by the residue theorem and then by
dimensional
regularization. Integrals involved in (\ref{eq:3.17})
are analogously  evaluated by first doing the 
$k_0$ integration and then evaluating the
$d^{n-1}k$ integral in dimensional
regularization.
We have verified that the integrals involved
(given below) in either case are finite in dimensional
regularization. These
integrals are\footnote{The integrals above generate several
(d-1)-dimensional 
integrals regularized dimensionally.Such odd-dimensional integrals
occur in statistical field 
theory in 3-dimensions and have been extensively, for example , in
Itzykson and Drouffe "Statistical field theory" Vol.I.;Cambridge
University Press, Cambridge, 1989.}
\begin{eqnarray}
\label{eq:listint1}
& & \int d^dk{(1,k_\mu,k_\mu k_\nu)\zeta^{0,1}
\over{[p+k)^2+i\epsilon](\vec
  k^2-i\epsilon)(\zeta+\zeta_0)^{0,1,2}}}
\delta(\zeta-{1\over2}\sqrt{-\epsilon\eta^2i})\nonumber\\
& &    \int d^dk{(1,k_\mu,k_\mu k_\nu)\zeta^{0,1}
\over{(k^2-i\epsilon)(\zeta+\zeta_0)^{0,1,2}}}
\delta(\zeta-{1\over2}\sqrt{-\epsilon\eta^2i})\nonumber\\
\end{eqnarray}
and those differing from the above by terms of
$O(\sqrt{\epsilon})$
in various factors in the denominator. We further need the
following integrals:
\begin{eqnarray}
\label{eq:listint2}
& & 
\int d^4k {\partial\over\partial\zeta}\biggl[{k_\mu
  k_\nu\over{k^2(k+p)^2}}
\biggr]\delta(\zeta);\ \int d^4k 
{\partial\over\partial\zeta}\biggl[{k_\mu\over{k^2}}\biggr]\delta(\zeta); 
\nonumber\\
& &
\int d^4k {\partial\over\partial\zeta}\biggl[{\eta\cdot(k+p)(k+p)_\mu
  k_\nu\over{k^2(k+p)^2}}
\biggr]\delta(\zeta)
\nonumber\\
\end{eqnarray}
(We also need integrals related to the above by a shift of
variable).
We evaluate these integrals for $p_0$ purely imaginary. We assume
the
analytic continuation of the divergent part
(which in our case is local) to real $p_0$ values.

Finally, we deal with the ghost loop
diagram in the Appendix B and show that it vanishes.

We, thus,see that the divergences in the gluon two point proper vertex in the one
 loop are identical to those with CPV prescription.

\section{Conclusions and Comments}

In this section, we summarize the results we have obtained. 
We started by emphasizing
that it is not required to interpose an interpretation of the $1/(\eta\cdot q)$
type of poles, and that the correct treatment of the axial gauge propagator
is obtained by its comparison with the Lorentz gauge path integral 
through FFBRS
transformations. It involves a well-defined contour integral and an effective
$\delta$-function term.The 2-point calculation presented here is a first
direct application of the propagator.
It illustrates and brings out several points:(i)
The calculation, which has no arbitrariness about it, 
is in no way more involved
than other prescriptions; (ii) It has only one vector $\eta$ already present
in the axial gauges required; thus limiting the number of the possible 
counterterms
in general; (iii) The usual expectations about the tadpole diagram in DR and
the ghost diagram are still valid in this formulation .With the calculations
so performed, we find, unlike the L-M /Uniform prescription, absence of any
nonlocal counterterms in the 2-point proper vertex.

It is possible that like the CPV prescription {[}2{]} that has one
vector needed,  this treatment, a one-vector treatment itself,
may also have only local counterterms
necessary for other n-point proper vertices and higher loop orders.

We, finally,comment on the possible approach 
to the study of locality of counter terms.It
is based on the result that correlates 
the axial Green's functions to the Lorentz
Green's functions that was obtained in references 13 and 7.The result {[}see
e.g.(46) of Ref.7{]} is valid for arbitrary Green's functions {[}as well as
operator Green's functions{]} .To illustrate the point, we state the result
for the 2-point function for which it reads:

\begin{eqnarray}
\label{eq:exres}
& & iG^{A\ \alpha\beta}_{\mu\nu}(x-y)=iG^{L\ \alpha\beta}_{\mu\nu}(x-y)
+i\int_0^1 d\kappa\int{\cal D}\phi e^{iS_{\rm eff}^M[\phi,\kappa]-i
\epsilon\int(A^2/2-{\bar c}c)d^4x}\nonumber\\
& & \times
\biggl(({\rm D}_\mu c)^\alpha(x)A^\beta_\nu(y)
+A^\alpha_\mu(x)
({\rm D}_\nu c)^\beta(y)\biggr)
\int d^4z{\bar c}(z)(\partial\cdot A^\gamma-\eta\cdot A^\gamma)(z)
\nonumber\\
& &
\end{eqnarray}

The above relation gives the value of the exact axial propagator 
\emph{compatible to the Green's functions in Lorentz gauges .}
The result is exact to all orders.As
mentioned in Ref.13, to any finite order in $g$,
the right hand side can be evaluated
by a finite sum of Feynman diagrams. These diagrams involve ,in particular,
ordinary propagators obtained from the mixed gauge effective action and a final
\( \kappa  \)-integral. A study of the form of the contributions to the last
term above should enable one to determine about the locality of counterterms.
This will be left for another work {[}17{]}.

\section*{Appendix A}
\setcounter{equation}{0}
\seceqaa

We shall take up diagram 1(b). We shall establish
the treatment for the graph in
terms of a contribution over
$C$ and an effective contribution. Up to overall factors, the
graphs is of the form
\begin{equation}
\label{eq:A1}
\int d^4k
V_{\mu\rho\lambda}(p,k,-k-p)V_{\nu\kappa\sigma}(-p,k+p,-k)D^{\sigma\rho}(k)D^{\lambda\kappa}(k+p),
\end{equation}
where
$D_{\sigma\rho}(k)$
is the exact propagator found in {]}7,8{]}. We
shall reduce the expression (\ref{eq:3.7}) in terms of the contour integral
and and an effective term. We do this for $\eta\cdot
p\neq0$. We chose  
$\epsilon$ such that $\sqrt{\epsilon}<<|\eta\cdot p|$. We
note that the expression for $D_{\sigma\rho}(k)$
consider with $D^0_{\sigma\rho}(k)$ except
  in the
small neighborhood ($\sim\alpha\sqrt{\epsilon},\
\alpha>>1$) of $\eta\cdot k=0$.Similarly, the
expression
for
$D_{\lambda\kappa}(k+p)$ coincide
with  $D_{\lambda\kappa}^0(k+p)$ except near
$\eta\cdot(k+p)=0$. For sufficiently small 
$\epsilon$, t he two regions  in the complex $k_0$-plane are
sufficiently  separated (separation $\Delta
k_0>>\sqrt{\epsilon}$).
Thus,, in the vicinity of
$\eta\cdot=0$,
$D_{\lambda\kappa}(k+p)\sim D^0_{\lambda\kappa}(k+p)$
and
vice versa. We note that  the vertex
factor
are
analytic in $k_0$. Thus, in the neighborhood
$|k_0-\vec\eta\cdot\vec k|\sim\alpha\sqrt{\epsilon}(\alpha>>1)$,
the remainder of the integrand (save the factor
$D_{\sigma\rho}(k)$ is analytic in $k_0$ and equals
$VVD^0_{\lambda\kappa}(k+p)$ up to $O(\epsilon)$. We can treat the
vicinity of the region
$\eta\cdot(k+p)=0$ in a
similar manner. We can thus express:
\begin{equation}
\label{eq:A2}
\int_{-\infty}^\infty dk_0 VVDD.= \int_{C_0} dk_0 VVDD
+\int_{C_1} dk_0 VVDD +\int_{C_2} dk_0 VVDD,
\end{equation}
where $C_0$ is the contour that runs mostly along the real $k_0$
axis
     together with two semicircular dips of reads
$\alpha\sqrt{\epsilon}$ around the points $\eta\cdot
k=0$ and
$\eta\cdot(k+p)=0$; $C_1$ and $C_2$ are compensating
closed contours as in {[}7{]}. On $C$, we can replace $D$ by
$D^0$ everywhere in both factors. On $C_1$, we can replace
$D(k+p)$ by $D^0(k+p)$ and on $C_2$, we can replace $D(k)$ by
$D^0(k)$. We can then extract the effective terms fro
$\int_{C_1}$ and $\int_{C_2}$
much as in {[}7{]}. During the process, 
we note that $VVD^0(k+p)$ is analytic in $k_0$ on $C_1$, etc.] Not
surprisingly,
we get:
\begin{eqnarray}
\label{eq:A3}
& & \int_{-\infty}^\infty dk_0 VVDD.= \int_{C_0} dk_0 VVD^0(k)D^0(k+p)
+\int dk_0
VVD^0(k+p)D^{\rm extra}\nonumber\\
& & +\int dk_0
VVD^0(k)D^{\rm extra},
\end{eqnarray}
where $D^{\rm extra}$ is given in (\ref{eq:Deff}). [Here we note the 
absence of a term involving $VVD^{\rm extra}D^{\rm extra}$ for
$\eta\cdot p\neq0$.]

We now discuss the evaluation of the first term on the right hand 
side side of (\ref{eq:A3}) and establish 
a simplified treatment. We note that a typical term
involved in this term is of the form:
\begin{equation}
\label{eq:A.4}
\int_{C_0} dk_0
{P(k,p)\over(k^2+i\epsilon)[(k+p)^2+i\epsilon]
\zeta^\alpha(\zeta+\zeta_0)^\beta},\ 0\leq\alpha,\beta\leq2
\end{equation}
where $\zeta=\eta\cdot k$ and
$\zeta_0=\eta\cdot p\neq 0$. Here $P(k,p)$ is a 
polynomial in $k$ and $p$. We can always partial fraction
${1\over\zeta^\alpha(\zeta+\zeta_0)^\beta},\
0\leq\alpha,\beta\leq2$ so that
each term has only single denominator of the the form
${1\over\zeta^\alpha}$ or
  ${1\over(\zeta+\zeta_0)^\beta}$
    ($1\leq\alpha,\beta\leq2$)
\footnote{We note that in the process of partial fractioning 
  we may generate coefficients of the type $\zeta^{-\gamma}$
  ($0\leq\gamma\leq2$) and 
these have to be kept track of in the final form.}

We thus have to evaluate integrals of the form
\begin{equation}
\label{eq:A.5}
\int_{C_0} dk_0
{P(k,p)\over(k^2+i\epsilon)[(k+p)^2+i\epsilon]\zeta^\alpha}
,\ 
0\leq\alpha\leq2
\end{equation}
and
\begin{equation}
\label{eq:A.6}
\int_{C_0} dk_0
{P(k,p)\over(k^2+i\epsilon)[(k+p)^2+i\epsilon](\zeta+\zeta_0)^\beta},\ 
0\leq\beta\leq2
\end{equation}
in addition to the integrals known for having only local
divergences . Note that in each of
the above integrals (\ref{eq:A.5}) and (\ref{eq:A.6}), the contour
$C$, now, can be chosen to have  only {\it one} semicircular
dip, t he
other one having become redundant, because, now, there are
poles {\it either} at $\eta=0$ {\it or} at
$\zeta+\zeta_0=0$. Furthermore, a change of
variable $k+p\rightarrow k$, casts the latter integral in 
the form (with $C$ suitably modified)
\begin{equation}
\label{eq:A.7}
\int_{C_0} dk_0
{P^\prime(k,p)\over(k^2+i\epsilon)[(k+p)^2+i\epsilon]\zeta^\beta},\ 
1\leq\beta\leq2
\end{equation}
which is of the same form as (\ref{eq:A.5}). 

We thus have to evaluate integrals of the 
form (\ref{eq:A.5}) with $C$ avoiding the pole at $\zeta=0$ 
by encircling it below at a radius    $\sim\alpha\sqrt{\epsilon}$.

These are known to be local {]}2{]}.

\section*{Appendix B}
\setcounter{equation}{0}
\seceqbb

In this appendix, we  treat the ghost diagrams that we
may not neglect for
$\lambda\neq0$. It 
reads up to overall factors
\begin{equation}
\label{eq:B.1}
I_G=\int d^4k
V_\mu(p,k,-k-p)V_\nu(-p,k+p,-k)D(k)D(k+p)
\end{equation}
We can evaluate $D(k)$ along the lines of {]}7{]},
{]}8{]} and establish an effective treatment of (\ref{eq:B.1}). We
have (for $\eta\cdot p\neq0$) 
\begin{eqnarray}
\label{eq:B.2}
& & 
I_G=\int d^3k\int_{C_0}dk_0 V_\mu V_\nu D^0(k)D^0(k+p)+\int
d^3k\int dk_0 V_\mu V_\nu(D^0(k)D^{\rm extra}(k+p)+\nonumber\\
& & 
D^{\rm
  extra}(k)D^0(k+p)).
\end{eqnarray}
We recall
\begin{equation}
\label{eq:B.3}
D^0(k)D^0(k+p)={1\over\eta\cdot
  k}{1\over\eta\cdot(k+p)}={1\over\eta\cdot
  p}\biggl[{1\over{\eta\cdot k}}
-{1\over{\eta\cdot(k+p)}}\biggr].
\end{equation}
Thus, we have to evaluate
\begin{equation}
\label{eq:B4}
\int d^3k\int_{C_0}{dk_0\over\eta\cdot k}V_\mu V_\nu,
  {\rm and}\ 
\int d^3k\int_{C_0}{dk_0\over\eta\cdot(k+p)} V_\mu V_\nu.
\end{equation}
We recall
\begin{equation}
\label{eq:B.5}
V_\mu(p,k,-k-p)=\eta_\mu+O(\epsilon),
\end{equation}
where the
$O(\epsilon)$ terms may arise from the $O(\epsilon)$ terms in 
${\rm S}_{\rm eff}$. In
each case, the $O(1)$ terms in the above integrals vanish by the use
of results similar to
those in (\ref{eq:3.4}). We shall assume that the
$O(\epsilon)$
terms exist in dimensional regularization and hence can be
ignored as $\epsilon\rightarrow0$.
Further, calculation along the lines of  {]}7{]} shows
\footnote{$A(k)$ is dimensionless and in fact turns out to be a constant.}
\begin{equation}
\label{eq:B.6}
D^{\rm extra}
=\delta\biggl(k^0-\vec\eta\cdot\vec
k-O(\sqrt{\epsilon})\biggr)
A(k),
\end{equation}
the second term in $I_G$ of (\ref{eq:B.2}) becomes:
\begin{eqnarray}
\label{eq:B.7}
& & \int d^3k\int dk_0 V_\mu V_\nu
D^0(k)\delta(k^0+p^0-\vec\eta\cdot(\vec k+\vec p)
+O(\sqrt{\epsilon}))A(k+p)\nonumber\\
& & =\int d^3 k {V_\mu V_\nu\over\eta\cdot
  k}A(k+p)|_{\eta\cdot k=-\eta\cdot p
  +O(\sqrt{\epsilon})}
\nonumber\\
& & ={1\over{\eta\cdot p+O(\sqrt{\epsilon})}}\int
d^3k V_\mu V_\nu A(p+k)|_{\eta\cdot k=-\eta\cdot p
  +O(\sqrt{\epsilon})}.
\end{eqnarray}
Again, we note that the $O(1)$ term above vanishes in
dimensional regularization. We shall assume that
the $O(\sqrt{\epsilon})$
term exists in dimensional regularization and hence  can
be ignored.

\section*{Acknowledgement}

The work in part was supported by grant for the DST project 
No. DST/PHY/19990170.

\clearpage
\begin{center}
   
    \begin{picture}(300,70)(165,55)
    \Gluon(330,49)(380,49)2 3  \Vertex(380,49)3
    \Gluon(380,49)(430,49)2 3
    \GlueArc(382,82)(28,-90,270)5 8 
 %   \GlueArc(395,82)(28,180,360)4 8
 \Text(382,32)[]{Fig 1(a)}    
\end{picture}

\vskip 1.0 true in

    \begin{picture}(300,70)(165,55)
    \Gluon(330,49)(384,49)2 3  \Vertex(384,49)3
    \GlueArc(415,49)(30,0,180)5 4 
    \GlueArc(415,49)(30,180,360)5 5 
    \Gluon(445,49)(495,49)2 3 \Vertex(445,49)3
    \Text(415,8)[]{Fig 1(b)}
  \end{picture}    
   \vskip 1.0 true in

\begin{picture}(300,100)(165,55)
\Gluon(330,49)(385,49)2 3  \Vertex(385,49)3
\Oval(415,49)(20,31)(0)
\Gluon(445,49)(495,49)2 3 \Vertex(445,49)3 
\Text(415,23)[]{Fig 1(c)}
\end{picture} \\

\end{center}
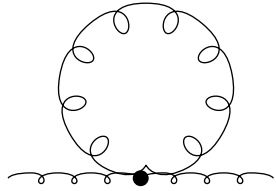
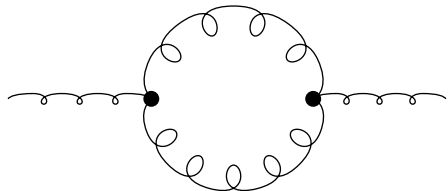
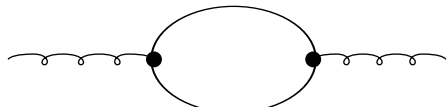
\end{document}